\documentclass[12pt,onecolumn]{article}
\usepackage{amssymb}
\usepackage{amsmath}
\usepackage[dvipdfmx]{graphicx}
\usepackage{cite}
\begin{document}





\begin{center}

{\large 
{\bf Renormalization group invariant of lepton Yukawa couplings 
} 
}

\vskip 0.3cm

Takanao Tsuyuki\\

\vskip 0.3cm

{ \it Graduate School of Science and Technology, Niigata University, Niigata 950-2181, Japan}

\begin{abstract} 
By using quark Yukawa matrices only, we can construct renormalization invariants that are exact at the one-loop level in the standard model. One of them $I^q$ is accidentally consistent with unity, even though quark masses are strongly hierarchical. We calculate a lepton version of the invariant $I^l$ for Dirac and Majorana neutrino cases and find that $I^l$ can also be close to unity. For the Dirac neutrino and inverted hierarchy case, if the lightest neutrino mass is 3.0 meV to 8.8 meV, an equality $I^q=I^l$ can be satisfied. These invariants are not changed even if new particles couple to the standard model particles, as long as those couplings are generation independent.
\end{abstract}

\end{center}

\section{Introduction}
\label{sec:intro}
In the standard model (SM) of particle physics, fermion masses are determined by their Yukawa couplings to the Higgs field. These couplings are free parameters and cannot be predicted within the SM. There may be some principles that determine the couplings at a high energy scale, such as a grand unified theory or a flavor symmetry. To compare the predictions that come from those high energy theories and observed values by experiments, we have to consider the renormalization group evolution of the parameters. Renormalization group invariants are useful because we can infer the high energy physics with low energy inputs. Invariants that are constructed only by quark Yukawa coupling matrices were reported in Ref. \cite{Harrison:2010mt}. One of them, which we call $I^q$, is close to unity. This fact seems surprising because their building blocks are far from one. This invariant is not changed even if there are some new particles that contribute to the beta functions in a generation-independent way since those contributions are canceled (as $\gamma_{u,d}$ in Ref. \cite{Harrison:2010mt}; see the Appendix). For example, if right-handed neutrinos exist and have Yukawa couplings of $O(1)$, they largely change the evolution of other Yukawa couplings \cite{Tsuyuki:2014xja}, but do not affect $I^q$.   

Today, we know that neutrinos have masses, so it is likely that they are coupled to the Higgs field as other fermions. There are still many mysteries in neutrinos: their type (Dirac or Majorana), the mass hierarchy (normal or inverted), and the absolute masses. We calculate a renormalization group invariant $I^l$ in the lepton sector that is similar to $I^q$ to see implications on these mysteries. If neutrinos are Dirac type, the relation between the neutrino Yukawa coupling matrix and their mass matrix is the same as other fermions. If neutrinos are Majorana type, we need some assumption to relate these matrices. In this paper, we consider simple cases of the type-$I$ seesaw mechanism \cite{Minkowski:1977sc,GellMann:1980vs,Yanagida:1979as}. To calculate $I^l$, we introduce a useful parametrization for neutrino Yukawa couplings.

As a result, we find that $I^l$ can also be $O(1)$ regardless of the type of neutrinos. 
 For the Dirac neutrino and the inverted hierarchy case, an equality between quark and lepton sector $I^q=I^l$ can hold if the lightest neutrino mass is 3.0 meV$\le m_{\rm lightest}\le$ 8.8 meV. For the Majorana neutrino cases, the invariant depends on many unknown parameters. 
Not only $I^q=I^l$, but also an interesting condition $I^l=1$ can be realized for Majorana cases. 

\section{Renormalization group invariants}
The Yukawa terms of the Lagrangian that we consider in this paper are
\begin{align}
-\mathcal{L}= \overline{u_R} Y_u \Phi^{c\dag} Q_L+\overline{d_R} Y_d \Phi^\dag Q_L+\overline{\nu_R} Y_\nu \Phi^{c\dag} L_L+\overline{l_R} Y_l \Phi^\dag L_L +{\rm H.c.}.
 \end{align}
We assume there are three right-handed neutrinos $\nu_R$. Renormalization group equations (RGEs) of Yukawa couplings \cite{Grzadkowski:1987tf,Antusch:2005gp} (see the Appendix) can be simplified by rewriting with Hermitian matrices $\alpha_f\equiv Y_f^\dag Y_f,\;(f=u,d,\nu,l)$. These quantities do not depend on the bases of right-handed particles $f_R$. There is a combination of the quark Yukawa matrices that is exactly invariant under the renormalization group evolution at the one-loop level \cite{Harrison:2010mt},
 \begin{align}
I^q&\equiv \left({\rm Det}[\alpha_u \alpha_d]\right)^{-2/3}\frac{{\rm Tr}[\alpha_u \alpha_d]}{{\rm Tr}[(\alpha_u \alpha_d)^{-1}]}. \label{eiq}
 \end{align}
This invariant does not depend on the basis of particles in the flavor space, since a unitary transformation of the basis of the left-handed quark doublet $Q_L$ is also canceled by taking traces and determinants. $I^q$ is invariant even if we introduce new particles, as long as they couple to the SM particle in a generation-independent way (see the Appendix).

In the basis that $\alpha_u$ is diagonal, we can express the Yukawa matrices as 
\begin{align}
\alpha_u &={\rm Diag}[m_u^2,\,m_c^2,\,m_t^2]/v^2,\\
\alpha_d&=V{\rm Diag}[m_d^2,\,m_s^2,\,m_b^2]V^\dag/v^2.
 \end{align}
$V$ denotes the quark mixing matrix and $v$ is the vacuum expectation value of the Higgs doublet $\Phi$. By substituting the observed values (renormalized to $M_Z$) \cite{Xing:2011aa,Charles:2004jd}, three factors in Eq. (\ref{eiq}) are
 \begin{align}
{\rm Det}[\alpha_u \alpha_d]&=(m_u m_c m_t m_d m_s  m_b)^2/v^{12}=6.26\times 10^{-36},\notag\\
{\rm Tr}[\alpha_u \alpha_d]&=\sum m_\beta^2m_i^2|V_{\beta i}|^2/v^4=2.63\times 10^{-4},\notag\\
{\rm Tr}[(\alpha_u \alpha_d)^{-1}]&=\sum m_\beta^{-2}m_i^{-2}|V_{\beta i}|^2v^4=5.76\times 10^{19}.\notag
 \end{align}
The summations are over $\beta=u,c,t$ and $i=d,s,b$.  These factors are much deviated from unity. The invariant is, however, close to unity,
\begin{align} 
I^q=1.35^{+0.54}_{-0.63}=0.72\dots 1.89
\end{align}
[1$\sigma$ errors are used in this paper except Eq. (\ref{ePlanck})]. The large errors of $I^q$ come from the errors of quark masses $\lesssim 30$\% \cite{Xing:2011aa}. It is surprising that this invariant is so close to and consistent with unity. By keeping leading terms and an approximation $V\simeq 1$, $I^q\simeq 1$ means 
\begin{align}
\frac{m_u m_t m_d m_b}{m_c^2 m_s^2}\simeq 1. \label{ei1}
\end{align}
Such a mass relation was pointed out in Ref. \cite{Davidson:1995ti}. In this paper, we calculate the invariant in the lepton sector,
 \begin{align}
I^l&\equiv \left({\rm Det}[\alpha_\nu \alpha_l]\right)^{-2/3}\frac{{\rm Tr}[\alpha_\nu \alpha_l]}{{\rm Tr}[(\alpha_\nu \alpha_l)^{-1}]}.
 \end{align}
We can see that $I^l$ is invariant in a manner discussed in Ref. \cite{Harrison:2010mt}. The relation between neutrino Yukawa couplings and the observed neutrino mass matrix depends on the type of neutrinos: Dirac or Majorana.

\section{Dirac neutrinos}

\begin{figure}[tb]
\begin{center}
\includegraphics[width=13cm]{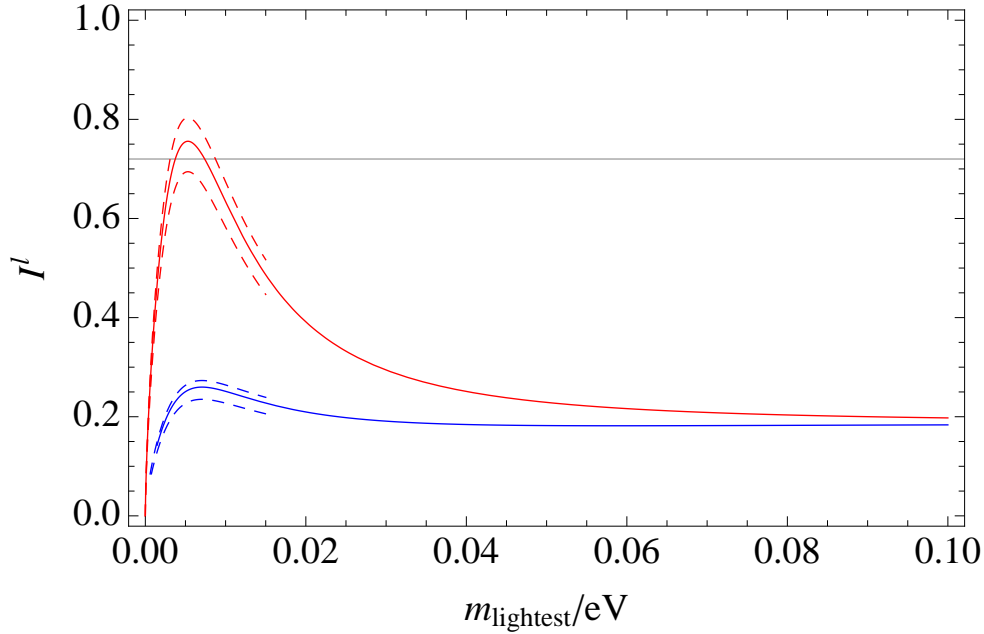}
\caption{Neutrino mass dependence of the renormalization group invariant $I^l$ for the Dirac neutrino case. Upper and lower curves show the IH and NH cases with 1$\sigma$ deviation roughly estimated from Eqs. (\ref{eilnhd}) and (\ref{eilihd}). The straight line shows the lower limit of $I^q$. }
\label{fD}
\end{center}
\end{figure}

If neutrinos are Dirac type, the neutrino Yukawa matrix $Y_\nu$ is directly converted to neutrino mass matrix $m_\nu$ as
\begin{align}
Y_\nu=\frac{1}{v}m_\nu=\frac{1}{v}U_RmU^\dag,\;m_{ij}\equiv m_i \delta_{ij}\;(i,j=1,2,3) \notag
 \end{align}
 $U$ and $U_R$ are unitary matrices that diagonalize $m_\nu$. In the basis where the charged lepton mass matrix is diagonal
 , $U$ corresponds to the neutrino mixing matrix. The invariant $I^l$ can be explicitly written similar to $I^q$,
\begin{align} 
I^l&=(m_1m_2m_3m_em_\mu m_\tau)^{-4/3}\frac{\sum m_\beta^2m_i^2|U_{\beta i}|^2}{\sum m_\beta^{-2}m_i^{-2}|U_{\beta i}|^2}. \label{eil}
\end{align}
The summations are over $\beta=e,\mu,\tau$ and $i=1,2,3$. Among the parameters in Eq. (\ref{eil}), absolute values of the neutrino masses cannot be determined by oscillation experiments. The strongest limit on the neutrino mass scale is set by the observation of the cosmic microwave background \cite{Ade:2013zuv},
\begin{align} 
\sum m_i < 0.23\;{\rm eV\; (95\% \;limit)}  \label{ePlanck}
\end{align}
 so we consider the region $m_i\le 0.1 \;\rm eV$ in this paper. The dependence of $I^l$ on the lightest neutrino mass is shown in Fig. \ref{fD}. We have used charged lepton masses (renormalized to $M_Z$ \cite{Xing:2011aa}) and the results of neutrino oscillation experiments \cite{Gonzalez-Garcia:2014bfa}. To better understand Eq. (\ref{eil}), we discuss three limiting cases below.

\begin{enumerate}
\item Degenerate masses \\
The simplest case is that neutrino masses are almost the same. If $m_1=0.1\rm\; eV$, by using mass differences given in Ref. \cite{Gonzalez-Garcia:2014bfa}, we obtain $m_2=0.1004$ eV and $m_3=0.11$ eV for the normal hierarchy (NH) case and $m_3=0.087$ eV for the inverted hierarchy (IH) case. By approximating $m_1\simeq m_2 \simeq m_3$ and noting that the charged lepton masses are hierarchical $m_e\ll m_\mu \ll m_\tau$, we can roughly estimate
\begin{align}
I^l&\simeq (m_1^3m_em_\mu m_\tau)^{-4/3}\frac{m_1^2 m_\tau^2\sum_i |U_{\tau i}|^2}{m_1^{-2}m_e^{-2}\sum_i |U_{e i}|^2} \notag \\
&= \left(\frac{m_em_\tau}{m_\mu^2 }\right)^{2/3}  \;(\equiv I^l_d) \notag \\
&=0.18648\pm0.00001.
 \end{align}
 Here we have used the unitarity of $U$. Interestingly, $I^l$ approximately does not depend on the neutrino masses, and this feature can be seen in the region of $m_1\sim 0.1$ eV of Fig. \ref{fD}.
\item{Normal Hierarchy (NH)} \\
We take $m_1\ll m_2 <m_3$. By subtracting the most dominant term of $I^l$,
\begin{align}
I^l\simeq \left(\frac{m_em_\tau}{m_\mu^2 }\frac{m_1 m_3}{m_2^2}\right)^{2/3}\frac{|U_{\tau 3}|^2}{|U_{e 1}|^2}= I_d^l\left(\frac{m_1 m_3}{m_2^2}\right)^{2/3}\frac{c_{23}^2}{c_{13}^2}. \label{eilnhd}
 \end{align}
We have used the conventional parametrization of $U$ given in Ref. \cite{Gonzalez-Garcia:2014bfa}. The experimental error of $I^l$ is roughly estimated from this equation to be $^{+5.1}_{-9.5}\%$. 
\item{Inverted Hierarchy (IH)} \\
We take $m_2 \gg m_3$. In this limit, $m_2=0.0495$ eV and $m_1=0.0487$ eV, so we can approximate $m_1\simeq m_2$. The dominant term of $I^q$ is
\begin{align}
I^l\simeq \left(\frac{m_em_\tau}{m_\mu^2 }\frac{m_3}{m_2}\right)^{2/3}\frac{|U_{\tau 1}|^2+|U_{\tau 2}|^2}{|U_{e 3}|^2}\simeq I_d^l\left(\frac{m_3}{m_2}\right)^{2/3}\frac{s_{23}^2}{s_{13}^2} \label{eilihd}
 \end{align} 
 Since $s_{13}\ll 1$, $I^l$ tends to be larger for the IH case than that for the NH case. The experimental error of $I^l$ is roughly estimated from this equation to be $^{+6.3}_{-8.1}\%$.
\end{enumerate} 

From Fig. \ref{fD}, we find that $I^l$ is $O(0.1)\sim O(1)$ for both mass hierarchy cases. In the IH case, $I^l=I^q$ can be satisfied for narrow and hierarchical regions of the lightest neutrino mass 3.0 meV$\le m_3 \le  8.8\;{\rm meV} \ll m_1,m_2$ (see Table \ref{tres}). This mass range gives 
\begin{align}
0.101\;{\rm eV}\le\sum m_i \le0.109\;\rm eV.
 \end{align}
Such a total neutrino mass is within the reach of future cosmological and astrophysical surveys   \cite{Abazajian:2011dt}. Unlike $I^q$, an interesting condition $I^l=1$ is highly excluded for both NH and IH cases.

\section{Majorana neutrinos}
If neutrinos are Majorana type, we need some assumption to relate their mass and the Yukawa matrix. Here we assume that the Majorana masses are generated through the type-I seesaw mechanism \cite{Minkowski:1977sc,GellMann:1980vs,Yanagida:1979as}. We add a Majorana mass term of three right-handed neutrinos to the \mbox{Lagrangian},
\begin{align}
\mathcal{L}\to \mathcal{L}-\frac{1}{2}\left(\overline{\nu_R}M\nu_R^c+\overline{\nu_R^c}M\nu_R\right),\;M_{IJ}=M_I\delta_{IJ}\;(I,J=1,2,3).
\end{align}
We take the basis of $\nu_R$ such that $M$ is diagonal. At a renormalization scale larger than the heaviest right-handed neutrino, RGEs of Yukawa couplings are the same as the Dirac neutrino case \cite{Antusch:2002rr}, so $I^l$ is invariant in this region. We expect that the Majorana masses of $\nu_R$ are much larger than the Dirac mass term. Then we obtain the neutrino mass matrix
\begin{align}
m_\nu = - v^2 Y_\nu^{T} M^{-1} Y_\nu. \label{eseesaw}
 \end{align}
In this case, we can parametrize the neutrino Yukawa couplings as  \cite{Casas:2001sr}
\begin{align}
\alpha_\nu=U \sqrt{m}R^\dag M R \sqrt{m}U^\dag /v^2. \label{eyy}
 \end{align}
$R$ is an arbitrary complex matrix that satisfies $R^{T}R=1$. This matrix cannot be determined by low energy experiments. The determinant of Eq.(\ref{eyy}) does not depend on $R$,
\begin{align}
{\rm Det}[\alpha_\nu]&={\rm Det}[m R^\dag M R/v^2]\notag\\
&=m_1m_2m_3 M_1 M_2 M_3/v^6.
 \end{align}
Here we have used Det[$R^\dag$]Det[$R$]$=(\pm 1)^2=1$. In the following, we assume degenerate masses $M_1=M_2=M_3\equiv M_N$. By these parametrization, the invariant is
\begin{align} 
I^l&= (m_1m_2m_3m_e^2m_\mu^2 m_\tau^2)^{-2/3}
\frac{{\rm Tr}\left[R^\dag R \sqrt{m} U^\dag m_l^2 U \sqrt{m}\right]}{{\rm Tr}\left[(R^\dag R \sqrt{m} U^\dag m_l^2 U \sqrt{m})^{-1}\right]}. \label{eilm}
\end{align}
$I^l$ does not depend on $M$ explicitly, but depend on $R^\dag R$. If $R$ is real, $R^\dag R=R^T R=1$. 
The unitary matrix $U$ have freedom of Majorana phases $\lambda_{1,3}$,
\begin{align}
U\equiv U^D D_M, \;D_M\equiv {\rm Diag}[e^{i\lambda_1},1,e^{i\lambda_3}],
 \end{align}  
where $U^D$ is the mixing matrix same as the Dirac neutrino case. Note that $D_M$ and $m$ are  commutative, so if $R$ is real, 
\begin{align}
\alpha_\nu=U^D D_M\sqrt{m}\sqrt{m}D_M^\dag U^{D\dag} /v^2=U^D m U^{D\dag} /v^2,
 \end{align}
which means that $I^l$ does not depend on the Majorana phases. 

The matrix $R^\dag R$ in Eq. (\ref{eilm}) is Hermitian and orthogonal. We introduce a useful parametrization for such a matrix,\footnote{
Another possible parametrization is $R^\dag R=\exp(iB)$, where $B$ is a real antisymmetric matrix \cite{Cirigliano:2007hb}. This is concise, but values of each component are not explicit.}
\begin{align} 
R^\dag R&=\sqrt{1+a^2}I-(\sqrt{1+a^2}-1)\frac{\vec{a}\vec{a}^T}{a^2}+iA,\\
A&\equiv \begin{pmatrix} 0&a_3&-a_2\\-a_3&0&a_1\\ a_2&-a_1&0 \end{pmatrix},\;\vec{a}^T\equiv (a_1,a_2,a_3),\; a\equiv |\vec{a}|.
\end{align}
If $R$ is real, $R^\dag R=R^T R=1$, which corresponds to $\vec{a}=\vec{0}$.\footnote{Without this condition, a general orthogonal Hermitian matrix can be written as 
\begin{align} 
\pm\left[\sqrt{1+a^2}I-(\sqrt{1+a^2}\pm1)\vec{a}\vec{a}^T/a^2+iA\right].
\end{align} 
}
 In principle, $a_i$ can take any real value, and $I^l$ can be infinitely large. Consider the case $a_2=a_3=0, a_1=a\gg 1$. In this limit, we obtain
\begin{align}
R^\dag R&\simeq a\begin{pmatrix} 0&0&0\\0&1&i\\0&-i&1\end{pmatrix}.
 \end{align}
Trace parts in Eq. (\ref{eilm}) do not depend on $m_1$, so the invariant is
\begin{align} 
I^l &\propto m_1^{-2/3}.
\end{align}
$I^l$ diverges in the limit $m_1\to 0$. The limit $a\gg 1$ is, however, highly unnatural, since it means that the absolute values of the components of $Y^\dag_\nu Y_\nu$ are much larger than that of $Y^T_\nu Y_\nu$ [compare Eq. (\ref{eseesaw}) and (\ref{eyy})]. To be concrete, we consider the region $|a_i|\le 5$.  Note that a small imaginary part of $R$ is enough to produce the baryon asymmetry of the Universe through leptogenesis \cite{Fukugita:1986hr,Tsuyuki:2014aia}.

\begin{figure}[tb]
\begin{center}
\includegraphics[width=13cm]{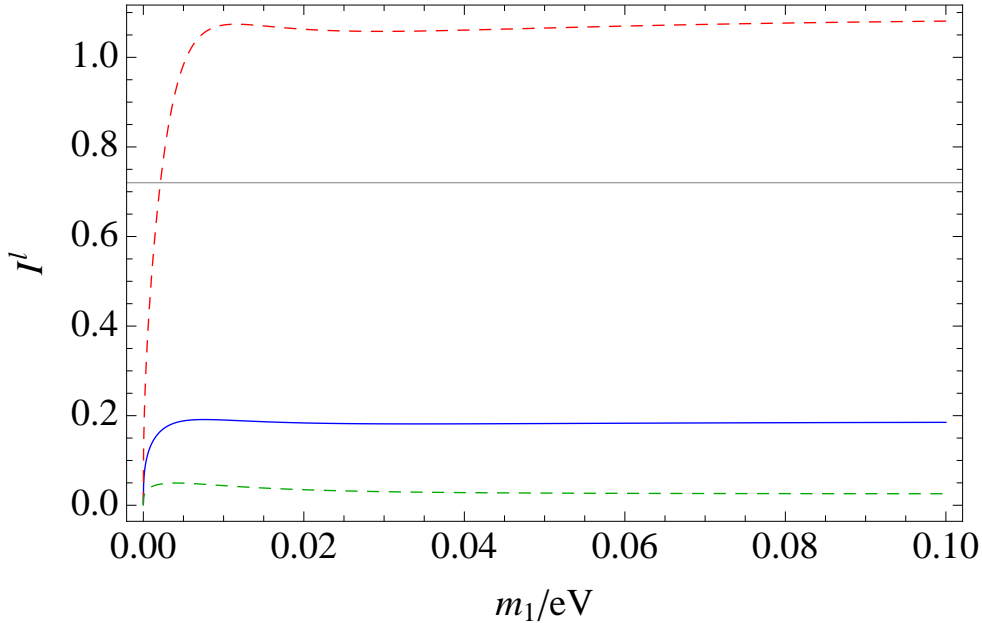}
\caption{Majorana neutrino, NH case. The blue solid curve is the case of real $R$, or $\vec{a}=0$. The upper and the lower dashed curves are cases in which $I^l$(0.1 eV) is largest and smallest [$\vec{a}^T=(5.00,2.89,0.572)$ and $\vec{a}^T=(-1.98,3.80,-5.00)$]. The lower limit of the invariant $I^q$ is also shown by a straight line.}
\label{fmnh}
\end{center}
\end{figure}

\begin{figure}[htb]
\begin{center}
\includegraphics[width=13cm]{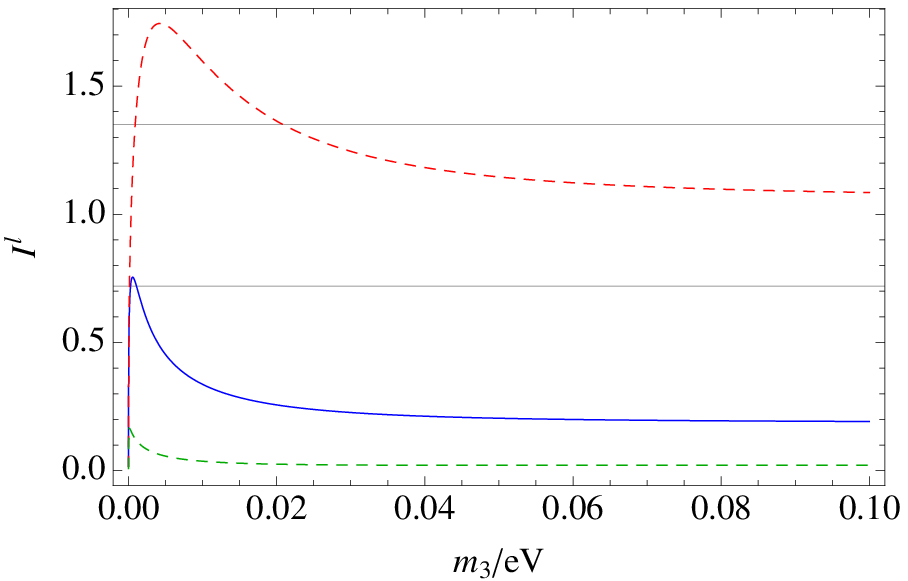}
\caption{Majorana neutrino, IH case. The upper and the lower dashed curves are taking $\vec{a}^T=(5.00,2.60,-0.235)$ and $\vec{a}^T=(-3.58,5.00,-5.00)$. Straight horizontal lines show the central value and lower limit of $I^q$ . Other notations are the same as in Fig. \ref{fmnh}.}
\label{fmih}
\end{center}
\end{figure}

The dependence of $I^l$ on the lightest neutrino mass is shown in Figs. \ref{fmnh} and \ref{fmih}. For each NH and IH case, we have shown three curves: (i) $a_i=0$, which corresponds to the case of real $R$, (ii) $I^l$(0.1 eV) is largest and (iii) smallest. Two Majorana phases are set to zero in the (ii) and (iii) cases. We have used charged lepton masses renormalized to $M_Z$ \cite{Xing:2011aa} and neutrino oscillation parameters\cite{Gonzalez-Garcia:2014bfa}. $I^l$ is defined at the energy scale larger than $M_N$, but we can expect that the renormalization group effect between $M_Z$ and $M_N$ is negligible. The reasons are as follows. First, $I^l$ include charged lepton masses in the form of ratios, which are almost constant in renormalization group evolution. Second, $I^l$ does not depend on the evolution of the overall scale of neutrino masses, and the renormalization effect on mixing angles is negligible \cite{Antusch:2002hy,Ohlsson:2013xva,Haba:2014uza} since the Yukawa coupling of the $\tau$ lepton is small in the SM (the effect is estimated to be smaller than $1^\circ$ \cite{Ohlsson:2013xva}). 

As we can see from Figs. \ref{fmnh} and \ref{fmih}, the invariant $I^l$ largely depends on the imaginary part of $R^\dag R$. If $R$ is real, the equality $I^l=I^q$ is  satisfied only in the IH case with small $m_3$,
\begin{align}
0.19\;{\rm meV}\le m_3 \le 1.5\;{\rm meV} \label{em3}
 \end{align}
 (see Table \ref{tres}). Majorana neutrino masses can be searched by not only cosmological observations but also neutrinoless double beta decay experiments that are sensitive to $m_{\beta\beta}\equiv |\sum U_{e i}^2 m_i|$. By Eq. (\ref{em3}), we obtain
\begin{align}
0.098\;{\rm eV}\le &\sum m_i \le0.100\;\rm eV,\\
0.018\;{\rm eV}\le &m_{\beta\beta} \le0.048\;\rm eV
 \end{align}
(Majorana phases are considered for $m_{\beta\beta}$). These regions can be searched in the foreseeable future \cite{Abazajian:2011dt,  GomezCadenas:2010gs}. If $R$ is complex, $I^l$ can be much larger, and an interesting condition $I^l=1$ can be satisfied for both NH and IH cases.

 \begin{table}[htb]
\caption{Summary of our numerical result. For Majorana cases, $R$ is assumed to be real. $I^l_{\rm peak}$ denote the extremum of $I^l$ (when the lightest neutrino mass is $m_{\rm lightest}=m_{\rm peak}$). $m_{I^l=I^q}$ is the lightest neutrino mass that satisfies the condition $I^l=I^q$ within one standard deviation.}
\begin{center}
\begin{tabular}{ccccc}\hline\hline
 &\multicolumn{2} {c}{Dirac}&\multicolumn{2} {c}{Majorana} \\
 & NH &IH&NH&IH \\ \hline 
$I^l_{\rm peak} $& $0.260^{+0.013}_{-0.025}$ & $0.756^{+0.047}_{-0.061}$ & $0.191_{-0.018}^{+0.010}$&$0.754^{+0.047}_{-0.061}$ \\
$m_{\rm peak}/\rm meV $& 7.06& 5.30 &7.59&0.568\\
$m_{I^l=I^q} /\rm meV$& -& 3.0\dots 8.8 &-&$0.19\dots 1.5$\\
\hline\hline
\end{tabular}
\end{center}
\label{tres}
\end{table}

\section{Summary}
\label{sec:conc}
We can construct renormalization group invariants with Yukawa coupling matrices only. 
As pointed out in Ref. \cite{Harrison:2010mt}, an invariant $I^q$ in the quark sector is close to one, even though quark masses are highly hierarchical. If this fact implies a rule for the structure of Yukawa matrices at the high energy scale, it would also hold in the lepton sector. In this paper, we have calculated an invariant $I^l$ of the lepton Yukawa couplings, which is defined similar to $I^q$. Note that $I^q$ and $I^l$ remain constant even if new particles couple to the SM particles generation independently. To calculate $I^l$ for Majorana neutrino cases, we introduced a useful parametrization of neutrino Yukawa couplings. We found that $I^l$ can also be $O(1)$ for both neutrino types. As the quark sector invariant, $I^l=1$ is consistent for Majorana neutrino cases, but not for Dirac neutrino cases. A hypothetical equality $I^l=I^q$ tends to favor inverted and strongly hierarchical neutrino masses, and such a mass spectrum can be explored by future experiments.

\section*{Acknowledgment}
I am grateful to Morimitsu Tanimoto for valuable comments.

\appendix

\section*{\appendixname\; Derivation of invariants}

In this Appendix, we see that $I^q$ is invariant in the SM and some extensions of it. The invariance of $I^l$ can be checked in a similar way. The RGEs of the quark Yukawa couplings are \cite{Grzadkowski:1987tf,Antusch:2005gp}
\begin{align}
Y_u^{-1}\frac{d}{dt}Y_u &=\frac{3}{2}(\alpha_u-\alpha_d)+\gamma_u I_{N_g\times N_g}, \label{eyuyu}\\
Y_d^{-1}\frac{d}{dt}Y_d &=\frac{3}{2}(\alpha_d-\alpha_u)+\gamma_d I_{N_g\times N_g}. \label{eydyd}
 \end{align}
$dt\equiv d\mu/(16\pi^2\mu)$, where $\mu$ is the renormalization scale. $I_{N_g\times N_g}$ is a unit matrix of the size $N_g$(= 3 in the SM) and $\gamma_{u,d}$ are flavor-independent terms of the beta functions. They come from the wave function renormalization of the Higgs doublet $\Phi$ and radiative corrections by gauge interactions. The RGEs of basis-independent quantities are calculated as
\begin{align}
\frac{d}{dt}{\rm Tr}[\alpha_u \alpha_d]&=2(\gamma_u+\gamma_d){\rm Tr}[\alpha_u \alpha_d],\\
\frac{d}{dt}{\rm Tr}[(\alpha_u \alpha_d)^{-1}]&=-2(\gamma_u+\gamma_d){\rm Tr}[(\alpha_u \alpha_d)^{-1}],\\
\frac{d}{dt}{\rm Det}[\alpha_u \alpha_d]&=2N_g(\gamma_u+\gamma_d){\rm Det}[\alpha_u \alpha_d].
 \end{align}
Then, the running of $I^q$ is
\begin{align} 
\frac{d}{dt}I^q&\equiv\frac{d}{dt}\left({\rm Det}[\alpha_u \alpha_d]^{-2/N_g}{\rm Tr}[\alpha_u \alpha_d]{\rm Tr}[(\alpha_u \alpha_d)^{-1}]^{-1}\right)\notag\\
&=\left( -\frac{2}{N_g}2N_g +2-(-2)\right)(\gamma_u+\gamma_d)I^q\notag\\
&=0,
\end{align}
so $I^q$ is a renormalization group invariant. We can easily see that ${\rm Tr}[\alpha_u \alpha_d]{\rm Tr}[(\alpha_u \alpha_d)^{-1}]$ is also invariant. Our derivation does not depend on the content of $\gamma_{u,d}$. It means that new particles that couple only to gauge fields or Higgs doublet do not change $I^q$. If new particles couple to quarks generation-independently, their contribution are also absorbed in $\gamma_{u,d}$, so $I^q$ remain invariant. For an example, right-handed neutrinos do not couple to quarks but couple to the Higgs doublet. They add ${\rm Tr}[\alpha_\nu]I_{N_g\times N_g}$ to the right-hand sides of Eqs. (\ref{eyuyu}) and (\ref{eydyd}). Such terms can largely change the evolution of $Y_u$ and $Y_d$ \cite{Tsuyuki:2014xja}, but $I^q$ remain invariant. 

The essential reason that we could find those invariants is that the generation-dependent terms can be canceled, $Y_u^{-1}\frac{d}{dt}Y_u+Y_d^{-1}\frac{d}{dt}Y_d\propto I_{N_g\times N_g}$. Two Higgs doublet models and minimal supersymmetric models change the Yukawa sector of the SM (see Ref. \cite{Antusch:2005gp} for RGEs), and then we cannot construct invariants in a parallel way.

{}

\end{document}